# Maximizing the thermoelectric performance of topological insulator $Bi_2Te_3$ films in the few-quintuple layer regime

Jinghua Liang[1], Long Cheng[1], Jie Zhang[1], Huijun Liu[1,*], Zhenyu Zhang[2,†]

[1] *Key Laboratory of Artificial Micro- and Nano-Structures of Ministry of Education and School of Physics and Technology, Wuhan University, Wuhan 430072, China*

[2] *International Center for Quantum Design of Functional Materials, Hefei National Laboratory for Physical Sciences at the Microscale, and Synergetic Innovation Center of Quantum Information and Quantum Physics, University of Science and Technology of China, Hefei, Anhui 230026, China*

Using first-principles calculations and Boltzmann theory, we explore the feasibility to maximize the thermoelectric figure of merit ($ZT$) of topological insulator $Bi_2Te_3$ films in the few-quintuple layer regime. We discover that the delicate competitions between the surface and bulk contributions, coupled with the overall quantum size effects, lead to a novel and generic non-monotonous dependence of $ZT$ on the film thickness. In particular, when the system crosses into the topologically non-trivial regime upon increasing the film thickness, the much longer surface relaxation time associated with the robust nature of the topological surface states results in a maximal $ZT$ value, which can be further optimized to ~2.0 under physically realistic conditions. We also reveal the appealing potential of bridging the long-standing $ZT$ asymmetry of *p*- and *n*-type $Bi_2Te_3$ systems.

---

[*] Author to whom correspondence should be addressed. Electronic mail: phlhj@whu.edu.cn
[†] Author to whom correspondence should be addressed. Electronic mail: zhangzy@ustc.edu.cn

**Introduction**

Fundamental research for clean and renewable energy source is a vitally important area of today's materials science. Among various promising energy processes, thermoelectric effects describe direct conversion of heat into electricity and vice versa. Despite many attractive features of thermoelectric devices, a major challenge in this field is searching for high-performance thermoelectric materials[1, 2] that could compete with traditional energy conversion methods. During the past decades, various innovative concepts and approaches have been developed to improve the efficiency of thermoelectric materials, such as low-dimensionalization[3, 4], energy filtering[5], band convergence[6, 7], and hierarchical architecturing[8]. These efforts have resulted in an overall better understanding of the underlying physical principles governing thermoelectric effects.

Intriguingly, theoretical predictions[9] and experimental studies[10, 11, 12] have shown that some of the good and well-established thermoelectric materials such as $Bi_2Te_3$, $Sb_2Te_3$, and $Bi_2Se_3$ also belong to a new class of quantum materials named three-dimensional (3D) topological insulators (TIs)[13, 14]. This novel quantum form of matter supports unique and topologically protected surface states[15, 16], and shares several essential materials features with good thermoelectric materials[17, 18]. For example, topological insulators usually contain heavy elements where the strong spin-orbit coupling effect can lead to band inversion in systems with small band gaps. Similarly, in order to obtain relatively larger power factor and lower lattice thermal conductivity, many good thermoelectric materials also have small band gaps and are consist of heavy elements. It is therefore natural to explore whether the robust topologically non-trivial surface states (TNSS) can be utilized to enhance their thermoelectric performance. Earlier attempts in addressing this fundamentally important question have resulted in highly controversial findings. For example, several theoretical studies of TI systems such as $Bi_2Te_3$ and $Bi_2Se_3$ concluded that the presence of the TNSS would suppress their thermoelectric figure of merit ($ZT$ values) [17, 19]. In contrast, the topological edge states in an idealized 2D TI model system have been exploited to yield large and anomalous Seebeck effects[18].

In this work, using first-principles calculations and Boltzmann transport theory, we provide systematic investigation of the effects of TNSS on the thermoelectric performance of $Bi_2Te_3$ films within the thickness range of $d = 1\sim6$ quintuple layers (QLs). The choice of the materials-specific $Bi_2Te_3$ system is based on the facts that it is the best thermoelectric material around room temperature, and is also a strong TI. Although it was previously shown that[20, 21, 22, 23] there is a large enhancement in the thermoelectric performance due to quantum confinement effect in the $Bi_2Te_3$ thin films (especially for the 1-QL film), such earlier studies did not explicitly take into account of the topological surface states. Here, we show that the *ZT* of the $Bi_2Te_3$ films exhibits a novel and generic non-monotonous dependence on the film thickness, peaked when the system enters the topologically non-trivial regime from the trivial one. Moreover, by fine tuning the relaxation time ratio between the bulk states and the TNSS within a physically realistic range, the *ZT* value can be optimized to ~2.0 at the critical thickness of $d = 3$ QLs. We also reveal that this approach can bridge the long-standing *ZT* asymmetry of *p*- and *n*-type $Bi_2Te_3$ systems[24, 25, 26], which may prove to be instrumental in future design of thermoelectric devices.

**Computational methods**

Our first-principles calculations were performed within the framework of density functional theory (DFT), as implemented in the Vienna *ab-initio* simulation package (VASP)[27]. The exchange-correlation energy is in the form of Perdew-Burke-Ernzerhof (PBE)[28] with generalized gradient approximation (GGA). The electron-core interaction is described by the projected augmented wave (PAW) method[29, 30, 31]. The spin-orbit coupling (SOC) is explicitly considered in our calculations since Bi and Te are both heavy atoms, and more importantly, the TI phase is driven by strong SOC. All the investigated $Bi_2Te_3$ QLs are modeled by using a hexagonal slab with a vacuum distance larger than 20 Å to avoid interactions between the QLs and its periodic images. The energy cutoff for plane wave expansion is set as 340 eV, and a $\Gamma$-centered $10\times10\times1$ *k*-point mesh is adopted for the Brillouin zone integration. In the calculations of transport properties, a fine $120\times120\times1$ *k*-mesh is used. The van der Waals (vdW) interaction between adjacent QLs is taken into account by means of the optB86b-vdW functional[32]. The relaxed atom positions are determined until the magnitude of the force acting on each atom is less than 0.05 eV/Å.

Based on the calculated band structures, the electronic transport properties can be determined by using the Boltzmann theory in the diffusive limit of transport, where the kernel is the so-called transport distribution function (TDF)[33]:

$$\Xi(\varepsilon) = \frac{1}{\Omega N_{\vec{k}}} \sum_{\vec{k}} \vec{v}_{\vec{k}} \vec{v}_{\vec{k}} \tau_{\vec{k}} \delta(\varepsilon - \varepsilon_{\vec{k}}), \qquad (1)$$

Here $\vec{v}_{\vec{k}}$ and $\tau_{\vec{k}}$ are the group velocity and relaxation time, respectively. $N_{\vec{k}}$ is the number of sampled $\vec{k}$ points in the first Brillouin zone. The volume $\Omega$ of the investigated QL system is defined as the product of the area of a unit cell and its thickness $d$. The in-plane Seebeck coefficient $S$, electrical conductivity $\sigma$, and electronic thermal conductivity $\kappa_e$ are then expressed as integrations of TDF:

$$S = \frac{ek_B}{\sigma} \int d\varepsilon (-\frac{\partial f_0}{\partial \varepsilon}) \Xi(\varepsilon) \frac{\varepsilon - \mu}{k_B T}, \qquad (2)$$

$$\sigma = e^2 \int d\varepsilon (-\frac{\partial f_0}{\partial \varepsilon}) \Xi(\varepsilon), \qquad (3)$$

$$\kappa_e = k_B^2 T \int d\varepsilon (-\frac{\partial f_0}{\partial \varepsilon}) \Xi(\varepsilon) \left[\frac{\varepsilon - \mu}{k_B T}\right]^2. \qquad (4)$$

where is $k_B$ the Boltzmann's constant, $f_0$ is the equilibrium Fermi function at the chemical potential $\mu$ and temperature $T$. Within the rigid-band picture, the chemical potential $\mu$ indicates the carrier concentration $n$ of the system, which is given by $n = \frac{1}{\Omega}(N - \int d\varepsilon g(\varepsilon) f_0(\mu, T))$. Here $g(\varepsilon)$ is the density of states, and $N$ is the number of valence electrons. It should be mentioned that the Lorenz number derived directly from Equations (3) and (4) is usually much larger than the metallic limit $2.4 \times 10^{-8}$ W$\Omega$K$^{-2}$ at low carrier concentration[34]. As an alternative, we adopt a modified Lorenz number of $1.5 \times 10^{-8}$ W$\Omega$K$^{-2}$, which is generally used for Bi-based bulk and low-dimensional thermoelectric materials[22, 24, 35, 36]. For the lattice thermal conductivity $\kappa_L$, we use the results obtained by molecular dynamics simulations[37], which is much faster than first-principles approach and can also give an accurate prediction of thermal conductivity by employing an appropriate interatomic potential and taking into account the size effect[37]. Detailed values of $\kappa_L$ used are given in Table S1. With all the transport coefficients available, the dimensionless figure of merit $ZT = S^2 \sigma T /(\kappa_e + \kappa_L)$ can be readily determined.

**Results and discussions**

The $Bi_2Te_3$ thin films can be viewed as stacking the QLs, each of which is given in the sequence of Te-Bi-Te-Bi-Te and are held together through vdW interaction (Figure 1(a)). Here we focus on the $Bi_2Te_3$ thin films in the few QLs regime. To understand the thermoelectric performance, we begin with a detailed analysis of their electronic properties. Figures 1(b)-(d) present the band structures of the $Bi_2Te_3$ films with thickness $d$ = 1~3 QLs (also see Figure S1 for those with $d$ = 4~6 QLs). Here the surface states are identified by the wave function projection method[38] using the following criteria: (1) for the 1- and 2-QL films, the critical percentages of the projections onto the top-two or the bottom-two *atomic layers* are 45% and 30%, respectively; (2) for the 3-, 4-, 5-, and 6-QL films, the critical percentages of the projections onto the top-most or the bottom-most *quintuple layer* are 60%, 50%, 40%, 40%, respectively. By counting the times that the surface states cross the Fermi level between time-reversal invariant momenta[39], we can readily confirm that there are only topologically trivial surface states in the 1- and 2-QL films, while there are topologically protected surface states in films with thicknesses $d \geq$ 3 QLs. Moreover, there exists an indirect band gap of 304 meV and 61 meV for the 1- and 2-QL films, respectively, while the thicker films ($d \geq$ 3 QLs) are metallic. Such different behaviors are due to strong interaction between the top and bottom surface states in the 1- and 2-QL films, which is suppressed in the thicker films ($d \geq$ 3 QLs) since the surface states decay exponentially beyond two QLs[38]. As indicated by the blue lines in the band structures of thicker films, there are bulk gaps that are spanned by the TNSS around the Fermi levels. The bulk gap value is 170 meV, 140 meV, 135 meV, 128 meV for the 3-, 4-, 5-, and 6-QL films, respectively. As discussed in the following, the existence of such bulk gaps has a strong influence on the thermoelectric properties of films with TNSS, where the relaxation times of the surface states inside and outside these gaps are drastically different.

As the bulk and surface states have been explicitly distinguished, we can calculate their respective contributions to the transport coefficients as bulk ($\sigma_{bulk}$, $\kappa_{e,bulk}$, $S_{bulk}$) and surface ($\sigma_{surf}$, $\kappa_{e,surf}$, $S_{surf}$) parts by inserting the respective TDF into Equation (2), (3) and (4). The total transport coefficients ($\sigma_{tot}$, $\kappa_{e,tot}$, $S_{tot}$) are then expressed as[40]:

$$\sigma_{tot} = \sigma_{bulk} + \sigma_{surf}, \tag{5}$$

$$\kappa_{e,tot} = \kappa_{e,bulk} + \kappa_{e,surf}, \tag{6}$$

$$S_{tot} = (\sigma_{bulk}S_{bulk} + \sigma_{surf}S_{surf})/\sigma_{tot}, \tag{7}$$

For the lattice thermal conductivity $\kappa_L$, however, we do not need to divide it into the surface or bulk parts since phonon is a collective excitation in condensed matter.

We first discuss the transport properties of the $Bi_2Te_3$ films with only trivial surface states ($d$ = 1~2 QLs). For these two films, the relaxation times of surface and bulk states are both assumed to be that of bulk $Bi_2Te_3$, which is 22 fs by fitting the experimental data[33]. Such treatment has been widely accepted in many nanoscale systems, such as Si nanowires[41], ZnO nanowires[42], and $Bi_2Te_3$ nanofilms[22, 23]. Since the bulk and surface states are assumed to have the same relaxation times for both 1- and 2-QL films, for simplicity, we only discuss the total transport coefficients and ZT values, which are plotted in Figure 2 as a function of the carrier concentration $n$ at room temperature. We see from Figure 2(a) that the electrical conductivities $\sigma$ of these two films are very small when the carrier concentration $n$ is smaller than $10^{19}$ cm$^{-3}$, and both increase quickly at larger $n$. Since the electronic thermal conductivity $\kappa_e$ is connected with the electrical conductivity $\sigma$ by Wiedemann-Franz law, it basically exhibits similar behavior and is thus not shown here. As for the Seebeck coefficients shown in Figure 2(b), there are two peaks at smaller carrier concentrations (especially for the 1-QL film), but their values vanish at larger $n$. These observations suggest that there should be a tradeoff between the electrical conductivities and the Seebeck coefficients such that the power factors $S^2\sigma$ and the *ZT* values (Figures 2(c) and 2(d)) can be optimized at a particular carrier concentration $n$. However, we should note that since the 1-QL film has a much larger band gap than the 2-QL film due to the quantum confinement effect, the Seebeck coefficient of the 1-QL film can acquire a higher absolute value than that of the 2-QL film. As a result, the *ZT* value of the 1-QL film is larger than that of the 2-QL film, and can be optimized to as high as 1.1 for the *p*-type carriers. Such a *ZT* is also larger than that of bulk $Bi_2Te_3$ in its pristine form (*ZT*=0.4)[43], which confirms that low-dimensionalization can indeed enhance the thermoelectric performance. On the other hand, we find that for these two films, the *p*-type systems are more favorable than the *n*-type systems in optimizing the *ZT* values. This inherent asymmetry is however undesirable for designing good thermoelectric devices that require *p*- and *n*-legs with comparative efficiencies.

We now focus on the $Bi_2Te_3$ films with topologically protected surface states (*d* = 3~6 QLs). Depending on their locations relative to the bulk gap, the surface states can have quite different relaxation times. Inside the bulk gap, the surface states are protected from backscattering, making the relaxation time of surface states much longer than that of bulk states. Outside the bulk gap, the backscattering of surface states is allowed due to the interaction between the surface and bulk states, resulting in the decrease of the relaxation time. As a consequence, we adopt the dual scattering time model[18], where the relaxation time of the surface states inside and outside the bulk gap are denoted as $\tau_1$ and $\tau_2$, respectively. As discussed above, $\tau_1 > \tau_2$ and $\tau_2$ is approximated to be the same as that of bulk states. By averaging the experimentally measured mean free path and Fermi velocity of the topological insulator $Bi_2Te_3$[44], the relaxation time of the TNSS ($\tau_1$) is estimated to be 550 fs, which is indeed much higher than the relaxation time of bulk states ($\tau_2 = 22$ fs), and thus a minimum relaxation time ratio ($r_\tau = \tau_1/\tau_2$) of 25 can be expected. Note that the relaxation time ratio $r_\tau$ is system-dependent, and in principle can be increased by introducing nonmagnetic impurities or disorders that tends to reduce $\tau_2$ without significantly changing $\tau_1$. For example, it can be deduced from experimental measurements[45,46] that $r_\tau$ of HgTe quantum-well can be increased to larger than 1000 by introducing disorders. For our investigated $Bi_2Te_3$ QL systems, the change of relaxation time ratio may be realized by random alloying[47], or by inducing environmental disorder[48] or stacking disorder[49]. One may also do this by isoelectronic substitution[50,51,52,53], or doping the system with nonmagnetic impurities such as Ca, Na, Sn, Ge[54,55,56,57], and etc. It is reasonable to expect that by appropriately controlling the composition and/or concentration of these nonmagnetic impurities or disorders, a certain value of relaxation time ratio can be realized.

To understand the influence of relaxation time ratio on the thermoelectric properties of $Bi_2Te_3$ films with TNSS, we plot in Figure 3 the room temperature transport coefficients and the *ZT* value of the *p*-type 3-QL film as a function of $r_\tau$, which ranges from 25 to 1000 according to the above discussions. For comparison, the contributions from the bulk and surface states are also shown. As done before for the 1- and 2-QL films, the carrier concentration $n$ of 3-QL film is fine tuned to optimize the total *ZT* value at each individual $r_\tau$ (see Figure S2). For example, at $r_\tau = 25$, the optimized carrier concentration is $1.8 \times 10^{19}$ cm$^{-3}$, while it is $3.7 \times 10^{19}$ cm$^{-3}$ at $r_\tau = 100$ (more data can be found in Table S2). As the surface states have a much larger relaxation time ($\tau_1 = 550$ fs and $\tau_2 \leq 22$ fs as $r_\tau \geq 25$), we see from Figure 3(a) that $\sigma_{surf} \gg \sigma_{bulk}$, which means that the surface states dominate the electronic transport. However, since $S_{surf}$ and $S_{bulk}$ have opposite signs, there is a competition between the contributions of the surface and bulk states[58, 59] according to Equation (7), but the former dominates because of its much larger electrical conductivity. It should be emphasized that the optimized chemical potential is always located around the edges of bulk gap, where the relaxation time of surface states ($\tau_1$ or $\tau_2$) will have a sudden change and affect the slope of TDF (see Figure S3~S6). According to Equation (2), electrons below and above bulk edges thus have quite different contributions and cannot be cancelled, leading to a $S_{surf}$ with anomalous sign (negative for *p*-type system) and high absolute value larger than 200 μV/K[18]. This picture is quite different from conventional metallic systems which exhibit very small Seebeck coefficients. We also want to mention that at a fixed carrier concentration, $S_{bulk}$ is independent of $r_\tau$ as the bulk states have the same relaxation time $\tau_2$ that is cancelled out in Equation (2), while $S_{surf}$ does not because the relaxation times of surface states are $\tau_1$ or $\tau_2$ depending on their energy relative to the bulk gap. Moreover, as surface states dominate the electronic transport at the optimized carrier concentration, the total Seebeck coefficient $S_{tot}$ have the same sign as $S_{surf}$. With the increasing of $r_\tau$, the contribution from the surface states becomes more and more pronounced, leading to a steady increase of the total power factor and *ZT* value (see Figures 3(c) and 3(d)). At relatively smaller $r_\tau$, the *ZT* value of such film with TNSS is actually smaller than the best of those with only trivial surface states (1.1 for 1-QL film, see Figure 2(d)). Nevertheless, the *ZT* value of the 3-QL film becomes higher if $r_\tau$ is tuned to be larger than a critical value of 55. This can be achieved by introducing nonmagnetic impurities or disorders into the system, as mentioned above. Further increase of $r_\tau$ will lead to even higher *ZT* value. In short, the very existence of the topologically protected surface states in the $Bi_2Te_3$ films provides a novel route to optimizing the thermoelectric performance by increasing the relaxation time ratio between the surface and bulk states.

Similar findings can be obtained for the *n*-type 3-QL film (see Figure S7). In particular, we find that the *ZT* values of the *n*-type system can be enhanced to be as large as those of the *p*-type system, which is not seen in the films with only trivial surface states ($d$ = 1~2 QLs). From an industrial viewpoint, *p*- and *n*-legs with similar mechanical and thermal properties are required for good thermoelectric devices. However, it usually has been much more difficult to enhance the thermoelectric performance of the *n*-type than the *p*-type $Bi_2Te_3$[24, 25, 26]. Here, as the TNSS contribute significantly to the thermoelectric performance, the electron-hole symmetry associated with the Dirac cone nature around the Fermi level ensures that the *ZT* values of the *p*- and *n*-type systems can be simultaneously optimized. By fine tuning the relaxation time ratio between the TNSS and bulk states, we thus provide an efficient way to bridge the long-standing *ZT* asymmetry of the *p*- and *n*-type $Bi_2Te_3$ systems.

We have also calculated the transport coefficients of thicker $Bi_2Te_3$ films ($d$ = 4~6 QLs), and found similar $r_\tau$ dependence as for the 3-QL film (see Figure S8~S13). As discussed above, a larger value of $r_\tau$ is desired to enhance the thermoelectric performance. Figure 4 plots the room-temperature *ZT* values for the *p*- and *n*- type $Bi_2Te_3$ films with thicknesses $d$ = 3~6 QLs at two typical $r_\tau$ values of 100 and 1000. For comparison, the results for films with only trivial surface states ($d$ = 1~2 QLs) are also given. For both small and large values of $r_\tau$, we see from Figures 4(a) and 4(b) that the *ZT* values exhibit a distinct non-monotonous dependence on the film thickness, which is qualitatively enriched from the generally monotonous behavior due to quantum size effects as predicted by Dresselhaus *et al*[3]. Such a striking new observation is generic in nature, and is inherently tied to the delicate competition between the surface and bulk contributions to the thermoelectric performance, coupled with the overall quantum size effects. We want to emphasize that such non-monotonous behavior is very robust, and it dose not depend on the exact value of the relaxation time ratio or the variation of lattice thermal conductivity (by introducing nonmagnetic impurities or disorders). We have tested that if we consider three additional relaxation time ratios ($r_\tau$ = 25, 50, and 75), or if an average thermal conductivity of 1.3 W/mK is used for all the investigated QL systems, the calculated *ZT* values remain exhibit a distinct non-monotonous dependence on the film thickness (see Figure S14 and S15). It should be mentioned that the jump in the *ZT* value between the topologically trivial and non-trivial regimes originates from their significantly different surface relaxation time, as discussed above. At sufficiently larger thicknesses, the contributions from the bulk states become more and more weighted, and the *ZT* values of such films approach that of bulk $Bi_2Te_3$. It should be noted that at the critical thickness (*d* = 3 QLs) where the TNSS start to appear, we observe almost identical *ZT* of about 2.0 for both the *p*- and *n*-type systems by using physically realistic $r_\tau$ value of 1000. Such *ZT* values are not only significantly larger than that of bulk $Bi_2Te_3$ and those of films with only trivial surface states, but also exceed most laboratory results reported so far. Moreover, if nonmagnetic impurities or disorders are introduced into the QL systems, there will be a possible reduction of the lattice thermal conductivity $\kappa_L$, and the corresponding *ZT* value can be further enhanced. As a test example, we show in Figure S16 that if $\kappa_L$ of each QL system is assumed to be reduced by half, the *ZT* values still show a robust non-monotonous behavior, and a maximal *ZT* of about 2.9 can be achieved at the critical thickness of 3 QLs.

To experimentally test the strong predictions presented in this work, one needs to first fabricate the atomically thin $Bi_2Te_3$ films, as already demonstrated by exfoliating a bulk sample to obtain suspended QL systems[60, 61]. Recently, a similar method has been used to synthesize freestanding $Bi_2Se_3$ films of only 1 QL, which in turn exhibits superior thermoelectric performance than the bulk counterparts[62]. We further note that a recent experimental work on the transport properties of $Bi_2Se_3$ thin film found that the surface transport diminishes abruptly below a critical thickness when the TI nature of the film is lost[63], which is very similar to our theoretical predictions. With the rapid developments of layer-resolved thermoelectric measurements[64, 65, 66], we anticipate that the present work will stimulate new experimental efforts and advances in this vitally important field.

**Conclusions**

In conclusion, our theoretical work offers some new insights into the inherent connections between the more traditional systems of thermoelectric materials and the new quantum materials of topological insulators. Beyond the well-known quantum size effect, we have shown that the thermoelectric performance of the $Bi_2Te_3$ films depends non-monotonously on the thickness, which is a generic property of such TI systems in the few-quintuple layer regime. As a consequence, a maximal *ZT* value of about 2.0 can be achieved when the system transits from the topologically trivial to non-trivial regime. Collectively, these findings could provide unique new design strategies to not only significantly enhance the thermoelectric performance of TI films, but also potentially bridge the long-standing *ZT* asymmetry of *p*- and *n*-type $Bi_2Te_3$ systems encountered by the broad thermoelectric research and industrial communities.

**Acknowledgment**

We thank Dr. Feng Liu for helpful discussions. We also acknowledge financial support from the National Natural Science Foundation (Grant Nos. 11574236, 51172167 and 1134006), and MOST of China (Grant Nos. 2013CB632502 and 2014CB921103).

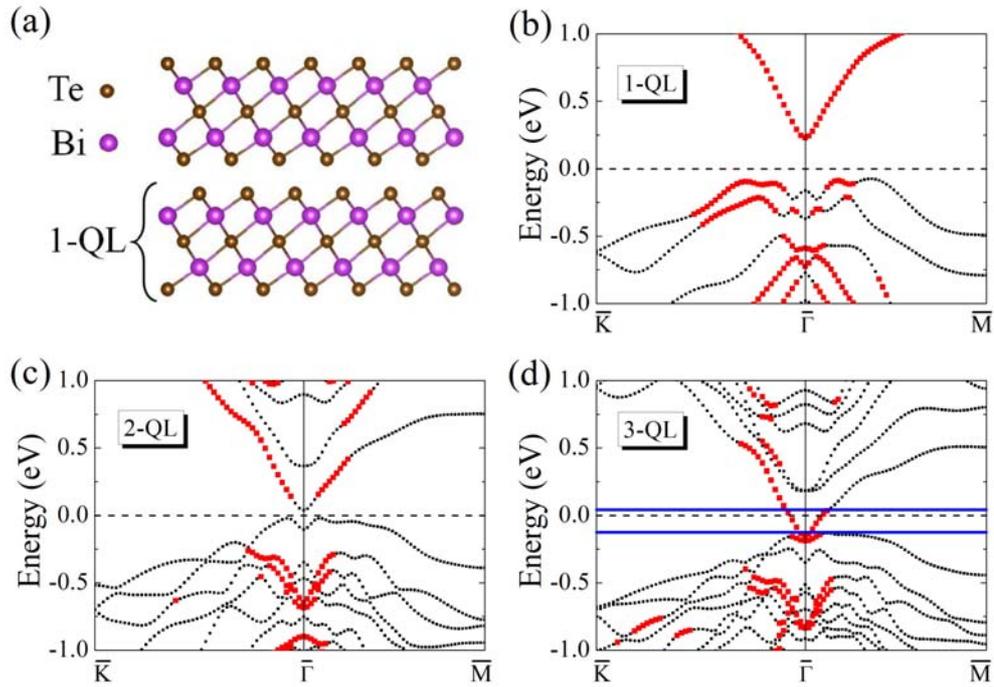

**Figure 1** (a) Side-view of $Bi_2Te_3$ thin films which are stacked by the QLs through vdW interaction. (b)-(d) Band structures of $Bi_2Te_3$ films with thicknesses $d$ = 1~3 QLs, where the surface states are marked by the red squares. The Fermi level is at 0 eV. The blue lines shown in (d) indicate the bulk gap.

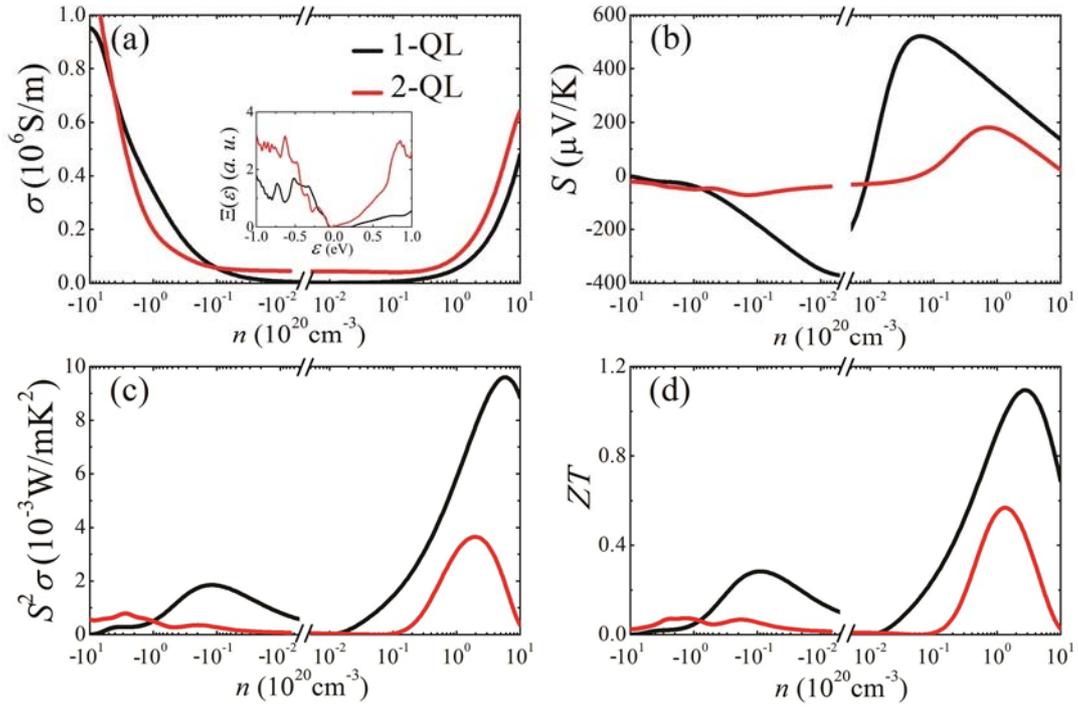

**Figure 2** Calculated (a) electrical conductivity $\sigma$, (b) Seebeck coefficient $S$, (c) power factor $S^2\sigma$, and (d) figure of merit $ZT$ as a function of the carrier concentration $n$ for $Bi_2Te_3$ films with thicknesses $d$ = 1~2 QLs. Positive and negative carrier concentrations represent *p*- and *n*-type carriers, respectively. The inset shows the corresponding TDF from which the transport coefficients are obtained.

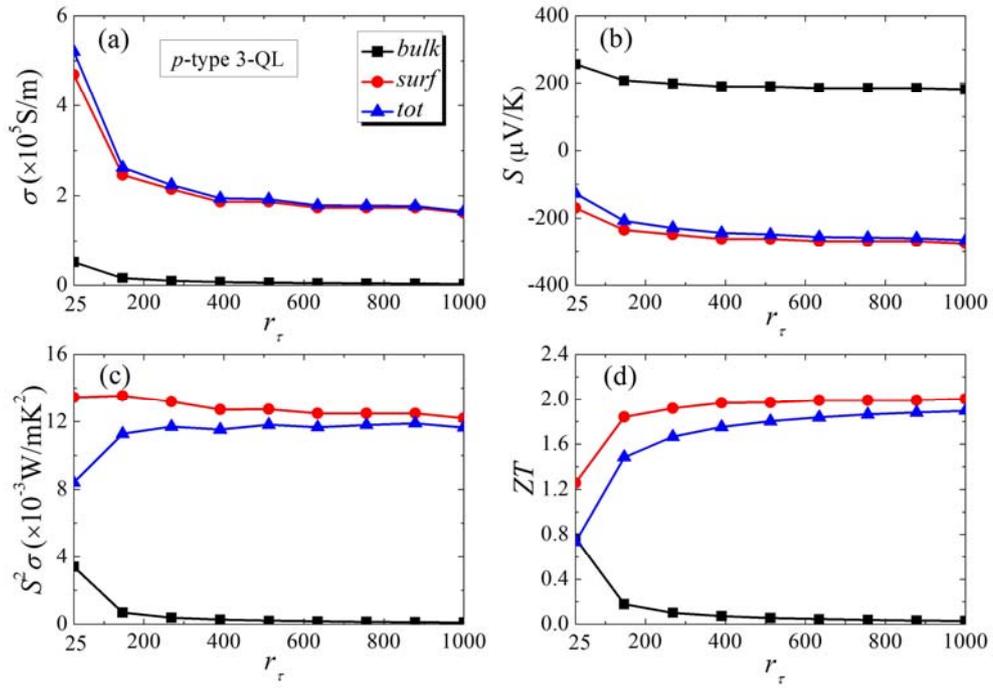

**Figure 3** Calculated (a) electrical conductivity $\sigma$, (b) Seebeck coefficient $S$, (c) power factor $S^2\sigma$, and (d) figure of merit $ZT$ of $p$-type 3-QL $Bi_2Te_3$ film as a function of the relaxation time ratio $r_\tau$. Contributions from the surface and bulk states are also shown.

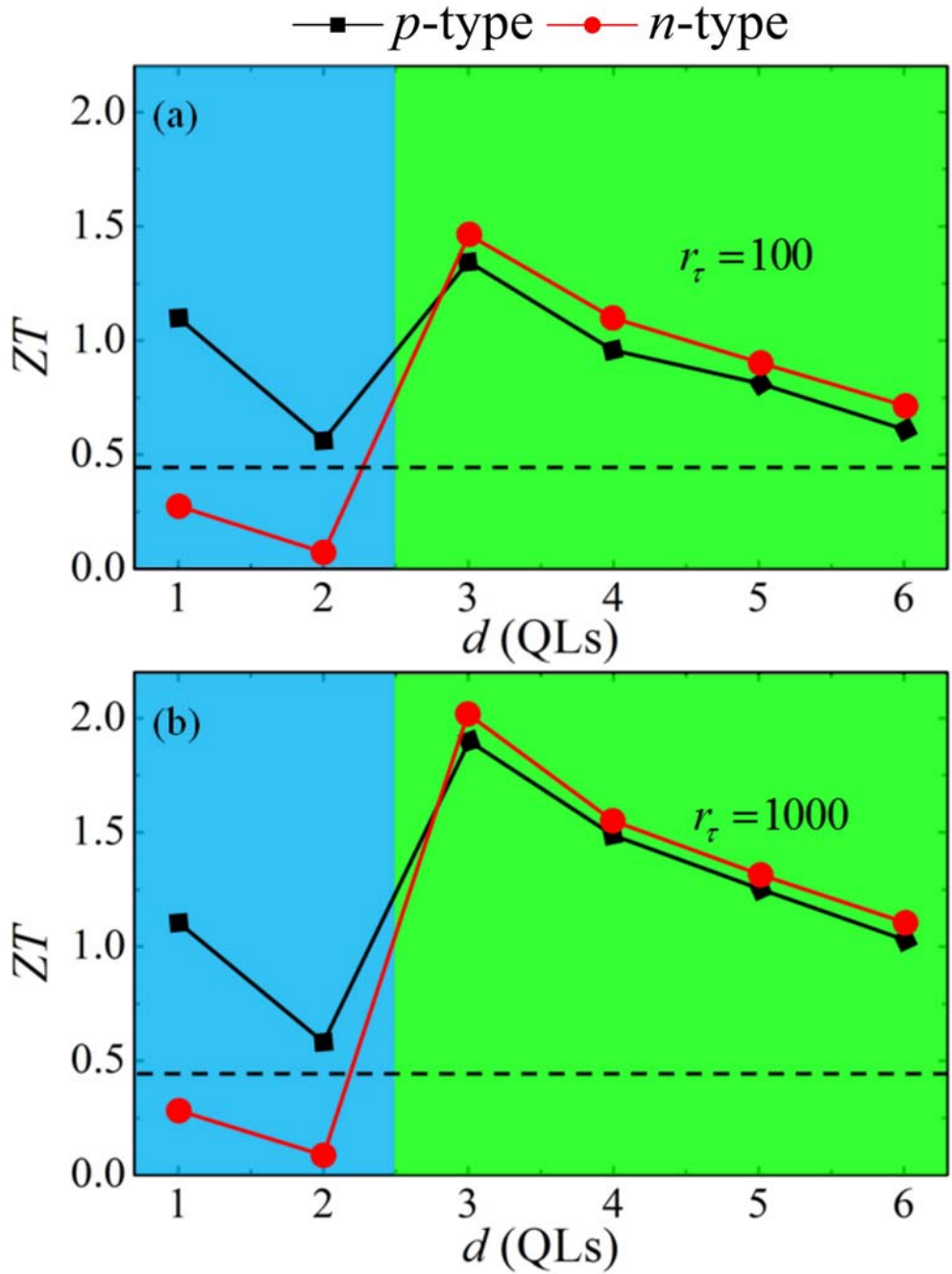

**Figure 4** Thickness-dependent *ZT* values of Bi$_2$Te$_3$ films. The blue and green regimes represent films with trivial and non-trivial surface states, respectively. The dashed lines indicate the *ZT* value of bulk Bi$_2$Te$_3$ in its pristine form. The relaxation time ratio in the topologically non-trivial regime of (a) and (b) is assumed to be 100 and 1000, respectively.